\documentclass[aps,prstab,reprint,amsmath,amssymb,groupedaddress]{revtex4-1}
\usepackage{hyperref}
\hypersetup{colorlinks=true, citecolor=blue, urlcolor=blue, linkcolor=blue}

\usepackage{graphicx}

\newcommand{\unit}[1]{\,\mathrm{#1}}
\newcommand{\funit}[2]{\,\mathrm{#1}/\mathrm{#2}}

\newcommand{\nexp}[1]{\times 10^{#1}}

\begin{document}

\title{Emittance preservation of an electron beam in a loaded quasi-linear plasma wakefield}

\author{Veronica K. Berglyd Olsen}
\email[]{v.k.b.olsen@cern.ch}

\author{Erik Adli}
\affiliation{University of Oslo, Oslo, Norway}

\author{Patric Muggli}
\affiliation{Max Planck Institute for Physics, Munich, Germany}
\affiliation{CERN, Geneva, Switzerland}

\date{\today}

\begin{abstract}
We investigate beam loading and emittance preservation for a high-charge electron beam being accelerated in quasi-linear
plasma wakefields driven by a short proton beam. The structure of the studied wakefields are similar to those of a long,
modulated proton beam, such as the AWAKE proton driver. We show that by properly choosing the electron beam parameters
and exploiting two well known effects, beam loading of the wakefield and full blow out of plasma electrons by the
accelerated beam, the electron beam can gain large amounts of energy with a narrow final energy spread (\%-level) and
without significant emittance growth.
\end{abstract}

\maketitle

\section{Introduction}\label{S:I}

Beam driven plasma wakefield accelerators~\cite{chen:1985} have the potential to offer compact linear accelerators with
high energy gradients, and have been of interest for several decades. With a relativistic charged particle drive beam
travelling through a plasma, a strong wakefield is excited that can be loaded by a trailing witness beam. When the
witness beam optimally loads the wakefield, an increase in absolute energy spread can be kept small. The concept has
been demonstrated experimentally in the past using electron drive beams accelerating electron witness
beams~\cite{rosenzweig:1988, blumenfeld:2007, kallos:2008, litos:2014}. 

A major challenge with plasma wakefield accelerators is, however, to accelerate a beam while keeping both energy spread
and emittance growth small. In the well described linear regime, valid when the beam density $n_{b}$ is much smaller
than the plasma density $n_{0}$, a non-linear transverse focusing force causes emittance growth of the witness beam.
The beam also sees a transversely and longitudinally varying accelerating field causing a spread in energy after the
beam has been accelerated~\cite{katsouleas:1987}. In the non-linear regime, where $n_{b} > n_{0}$, a bubble is formed by
the transverse oscillations of the plasma electrons, gathering in a sheath around an evacuated area filled with only
ions. The ions, assumed stationary, form a uniform density ion channel creating a focusing force that varies linearly
with radius. This focusing force preserves emittance~\cite{rosenzweig:1991}.

In this paper we present simulation results showing how emittance preservation of a high charge density witness beam can
be ensured when accelerated by a proton drive beam producing quasi-linear wakefields~\cite{rosenzweig:2010}. By
quasi-linear wakefields we here mean wakefields with only partial blow out of the plasma electrons in the accelerating
structure (bubble). The key idea is to have enough charge in the witness beam to at the same time load the wakefield to
produce low relative energy spread, and completely blow out the electrons left in the accelerating structure after the
beam to reach conditions that preserves emittance. The results have importance for the preparation of AWAKE
Run~2~\cite{adli:2016}, and possibly other applications in the quasi-linear regime.

\subsection{AWAKE Run 2}\label{S:I:AWAKE}

\begin{figure}[hbt]
    \includegraphics[width=0.99\linewidth,trim={1mm 2mm 1mm 2mm},clip]{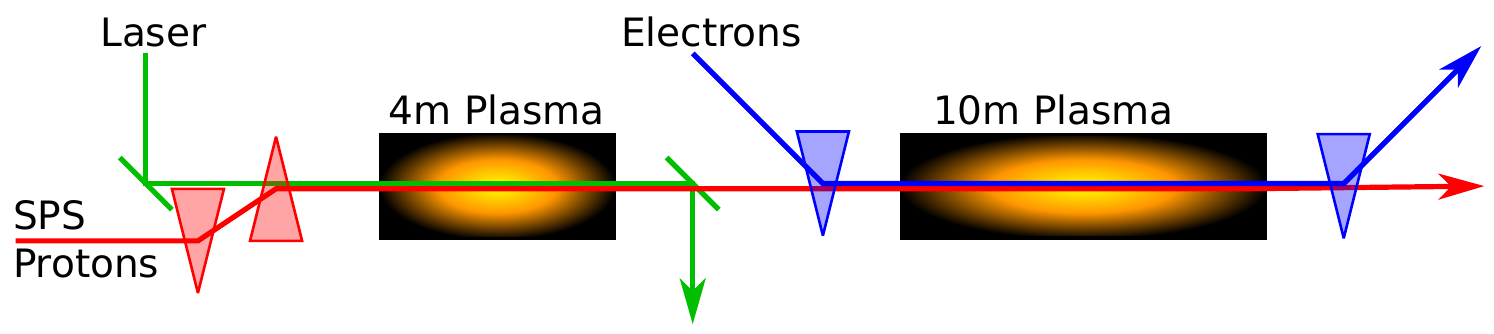}
    \caption{\label{Fig:AWAKER2} A simplified illustration of the experimental setup for AWAKE Run 2. The SPS proton
        beam undergoes self-modulation in the first plasma section. The electron witness beam is injected into the
        accelerating structure, and undergoes acceleration in the second plasma
        section~\cite{berglyd_olsen:2015, adli:2016}}.
\end{figure}

The energy carried by electron drive beams used in previous plasma wakefield experiments have been on the order of
$100\unit{J}$ and the propagation length typically no more than $1\unit{m}$~\cite{blumenfeld:2007, caldwell:2009}.
For high-energy physics application a higher total beam energy is often desired. For instance, the energy of a
high-charge electron beam accelerated to $1\unit{TeV}$ with $1\nexp{10}$ electrons, similar to the beam that could be
produced by the International Linear Collider, is $1.6\unit{kJ}$. Using electron beams as drivers, a large number of
plasma stages is required to reach an energy of a $\unit{kJ}$ for the accelerated beam. However, staging plasma
accelerators without reducing the effective gradient and spoiling the beam quality is challenging
\cite{steinke:2016, lindstrom:2016}.

Proton beams available at CERN carry significantly more energy than available electron beams, $19\unit{kJ}$ for the SPS
beam~\cite{gschwendtner:2016}, allowing for much longer plasma wakefield accelerator stages. The SPS beam is orders of
magnitude longer than the plasma wavelengths needed for such applications, and it does not drive a strong
wake~\cite{gschwendtner:2016}. By letting the proton beam undergo self-modulation before injecting the witness beam into
the accelerating structure, stronger wakefields are excited. The self-modulation is produced by the transverse fields
generated by the beam acting upon itself, causing regions of the beam to rapidly defocus~\cite{kumar:2010}. The
modulation frequency is close to that of the plasma electrons, and produces a train of short proton bunches along the
beam axis with a surrounding halo of defocused particles. This train of bunches resonantly drives wakefields to large
amplitudes.

AWAKE at CERN is a proof of concept proton beam driven plasma wakefield accelerator experiment
\cite{awake_collaboration:2014}, currently in its first phase of operation. The experiment uses a $400\unit{GeV}$ proton
beam delivered by the SPS as its driver, and a single $10\unit{m}$ plasma stage with a nominal plasma density of
$7\nexp{14}\unit{cm}^{-3}$~\cite{gschwendtner:2016}. This plasma density corresponds to $\lambda_{pe} = 1.26\unit{mm}$
and is matched to the transverse size of the proton beam such that $k_{pe}\sigma_{x,y,pb} = 1$~\cite{lu:2005}, where
$k_{pe} = 2\pi/\lambda_{pe}$ is the plasma wave number, $\lambda_{pe} = 2\pi c/\omega_{pe}$, and
$\omega_{pe} = \left(n_0e^2/m_e\epsilon_0\right)^{1/2}$ is the plasma electron angular frequency.

The aim of the first phase of the experiment is to demonstrate self-modulation of the proton beam. The aim of the second
phase, in 2018, is to sample the wakefield with a long electron beam ($\simeq\lambda_{pe}$). The study presented here
has relevance for Run~2~\cite{adli:2016}, starting in 2021 after the LHC long shutdown 2, and aims to demonstrate
acceleration of a short electron beam ($\ll\lambda_{pe}$) to high energy and with a minimal increase in emittance and
absolute energy spread.

The plans for AWAKE Run~2 propose to use two plasma sections, as illustrated in Fig.~\ref{Fig:AWAKER2}. The first
section of about $4\unit{m}$ is the self-modulation stage where the proton beam undergoes self-modulation without the
electron beam present. The electron witness beam is then injected into the modulated proton beam before section two
where it undergoes acceleration. The self-modulated proton beam does not produce a fully non-linear wakefield, and
therefore not all plasma electrons are evacuated from the plasma bubble. The result is that the focusing force does not
increase linearly with radius and the accelerated beam emittance is not preserved. 

\section[\label{S:M}]{Method}

The focus in this study is on the beam loading of the wakefields driven by the proton beam. Studies of self-modulated
proton beams show that the beam evolves as it propagates through a uniform plasma~\cite{lotov:2011}, but small
variations in the plasma profile the modulation, and thus the wakefields, may be stabilised over long
distances~\cite{lotov:2011, lotov:2015, caldwell:2011}. To study the witness beam evolution in a stable wake,
independent of the dynamics of the self-modulation, we use a single, non-evolving proton bunch as driver. The proton
beam parameters are chosen so that key features of the wake \textendash the plasma electron density in the wake and the
longitudinal electric field \textendash are the same as in the wake of a self-modulated proton beam with AWAKE baseline
parameters~\cite{gschwendtner:2016}. Both the proton beam and the witness beam have Gaussian longitudinal charge
distribution and bi-Gaussian transverse charge distributions.

We have previously studied the beam loading in a proton beam wake using the full particle-in-cell (PIC) code
OSIRIS~\cite{fonseca:2002} with 2D cylindrical-symmetric simulations. The studies~\cite{berglyd_olsen:2015,
berglyd_olsen:2016} primarily looked at beam loading, energy gain and energy spread, as well as different approaches to
creating a stable drive beam structure based on previous self-modulation studies. In order to study the witness beam
emittance evolution we use the recently released open-source version of QuickPIC~\cite{huang:2006, an:2013}. QuickPIC is
a fully relativistic 3D quasi-static PIC code. It does not suffer from the numerical Cherenkov effect that full PIC
codes do~\cite{godfrey:1974,lehe:2013}, making it a well suited tool to study emittance preservation. All simulation
results in this paper were obtained using QuickPIC open-source~\cite{quickpic:web}.

\begin{figure}[hbt]
    \includegraphics[width=\linewidth,trim={2mm 0mm 2mm 0mm},clip]{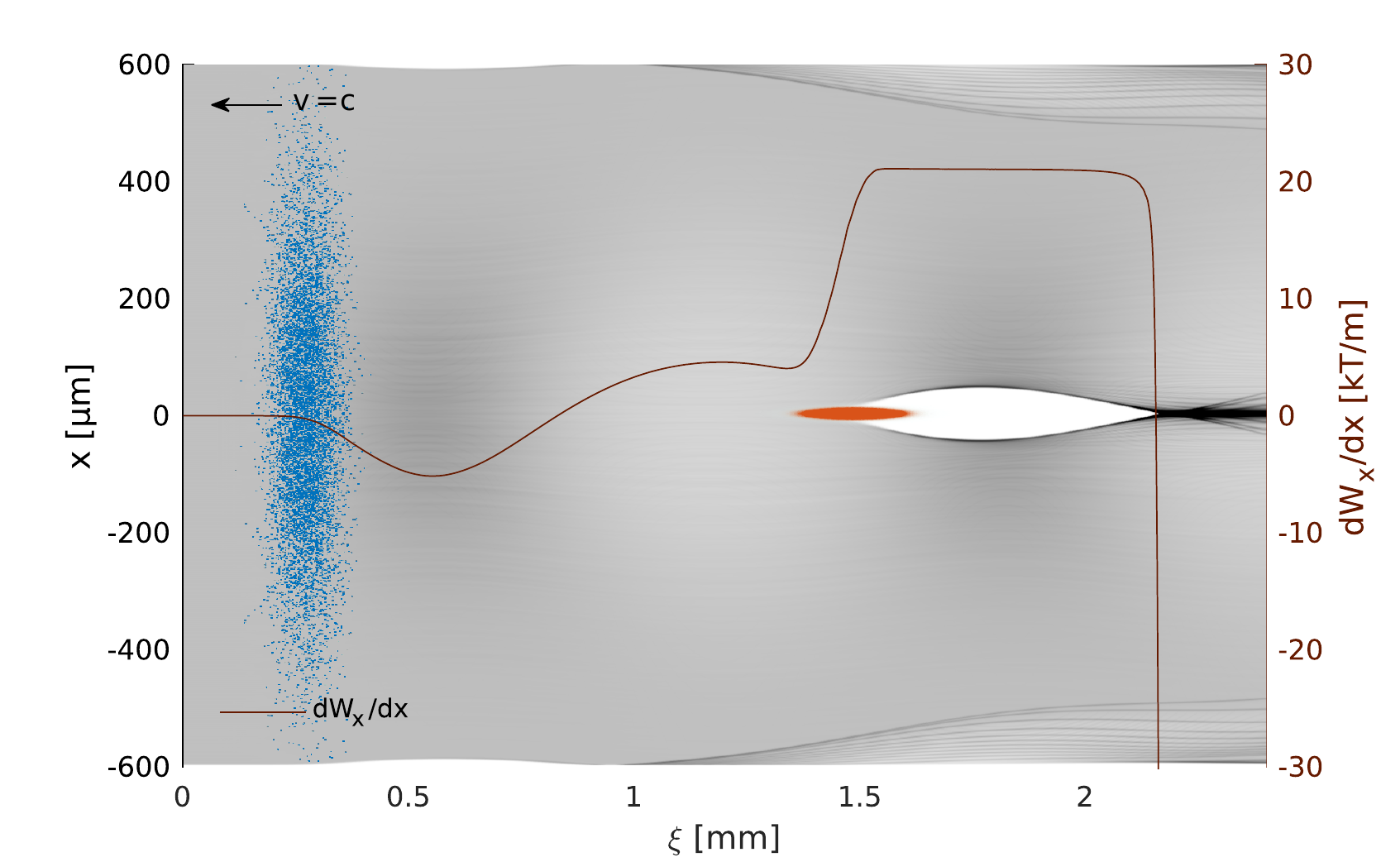}
    \caption{\label{Fig:PlasmaDenTWake} QuickPIC simulation results showing the initial time step for the single proton
        drive beam and witness beam setup. Plasma electron density is shown in grey with the drive beam (blue) and the
        witness beam (red) superimposed. The line plot indicates the transverse wakefield gradient
        $\textrm{d}W_{x}/\textrm{d}x$ where $W_{x} = E_{x} - v_{b} B_{y}$, evaluated along the beam axis. Beams move to
        the left.}
\end{figure}

\subsection{Drive Beam Parameters}\label{S:M:Setup}

The modulation process used in AWAKE does not reach the fully non-linear regime and thus does not produce a bubble void
of plasma electrons. When the SPS beam, containing $3\nexp{11}$ protons~\cite{gschwendtner:2016}, enters the second
plasma section (Fig.~\ref{Fig:AWAKER2}), the peak electric field is expected to be about $500\funit{MV}{m}$. The
plasma electron density is only depleted to around $65\%$ of nominal value at the point where we inject the
electron beam~\cite{awake_collaboration:2016}. The plasma electron density depletion and the peak field are replicated
using a single bunch with $1.46\nexp{10}$ protons ($2.34\unit{nC}$), a length $\sigma_{z} = 40\unit{\mu m}$
($7\unit{kA}$), and a transverse size $\sigma_{x,y,pb} = 200\unit{\mu m}$. The beam peak density is $0.83\cdot n_{0}$
and results in a quasi-linear wake. To avoid transverse evolution of the proton driver, emulating the stable propagation
of the self-modulated beam~\cite{lotov:2011, lotov:2015, caldwell:2011}, we freeze the transverse evolution of the
equivalent proton bunch by increasing the particles mass by six orders of magnitude.

\subsection{Witness Beam Parameters}\label{S:M:Setup}

In order to prevent large amplitude oscillations of the witness beam particles, which may cause additional energy spread
as well as emittance growth, we consider a witness beam matched to the plasma density. The matched beam transverse
size~\cite{esarey:1996} is:
\begin{equation}
    \sigma_{x,y,eb}=\left(\frac{2c^{2}\epsilon_{\mathrm{N}}^{2}m_{e}\varepsilon_{0}}{n_{pe}e^{2}\gamma}\right)^{1/4}.
    \label{EQ:Matched}
\end{equation}

We assume an initial normalized emittance of $\epsilon_{\mathrm{N}} = 2\unit{\mu m}$. This emittance is possible to
produce with a standard RF-injector, while at the same time yielding a sufficiently narrow beam.

Beam loading by a short witness beam is sensitive to its position relative to the electric field~\cite{tzoufras:2009}
as well as, at low energy, to its de-phasing with respect to the wakefields. To eliminate de-phasing of the witness
beam, the initial beam energy is set such that $\gamma_{eb} = \gamma_{pb} = 426.3$, giving an energy of $217\unit{MeV}$.
A lower initial energy is likely to be sufficient for AWAKE Run~2 injection.

Equation~(\ref{EQ:Matched}) yields a transverse size $\sigma_{x,y,eb}$ of $5.25\unit{\mu m}$, which is narrow compared
to the drive beam $\sigma_{x,y,pb} = 200\unit{\mu m}$. The bunch length was set to $\sigma_{z} = 60\unit{\mu m}$ based
on earlier beam loading studies~\cite{berglyd_olsen:2016}. The charge is adjusted to $100\unit{pC}$ for optimal beam
loading, as discussed in the next section. We refer to the defined drive beam and witness beam parameter set as the
\emph{base case}. Figure~\ref{Fig:PlasmaDenTWake} shows the two beams \textendash the proton beam in blue, the trailing
electron beam in red, and the plasma electron density in grey \textendash from a QuickPIC simulation of the initial time
step, for the base case parameters.

\subsection{Simulation parameters}\label{SIM}

The relatively small size of the witness beam puts constraints on the transverse grid cell size and number in the
simulations. We need a small size to resolve the narrow electron beam, and a large number of grid cells to resolve the
much wider proton beam and its wakefields. We use a transverse grid cell size of $1.17\unit{\mu m}$, and of
$2.34\unit{\mu m}$ for the longitudinal grid cells for the simulations presented in section~\ref{S:BL}. The witness beam
was simulated with $16.8\nexp{6}$ and the drive beam with $2.1\nexp{6}$ non-weighted particles, and the plasma electrons
with $1024 \times 1024$ weighted particles per transverse slice.

\begin{figure}[hbt]
    \includegraphics[width=\linewidth,trim={2mm 0mm 2mm 0mm},clip]{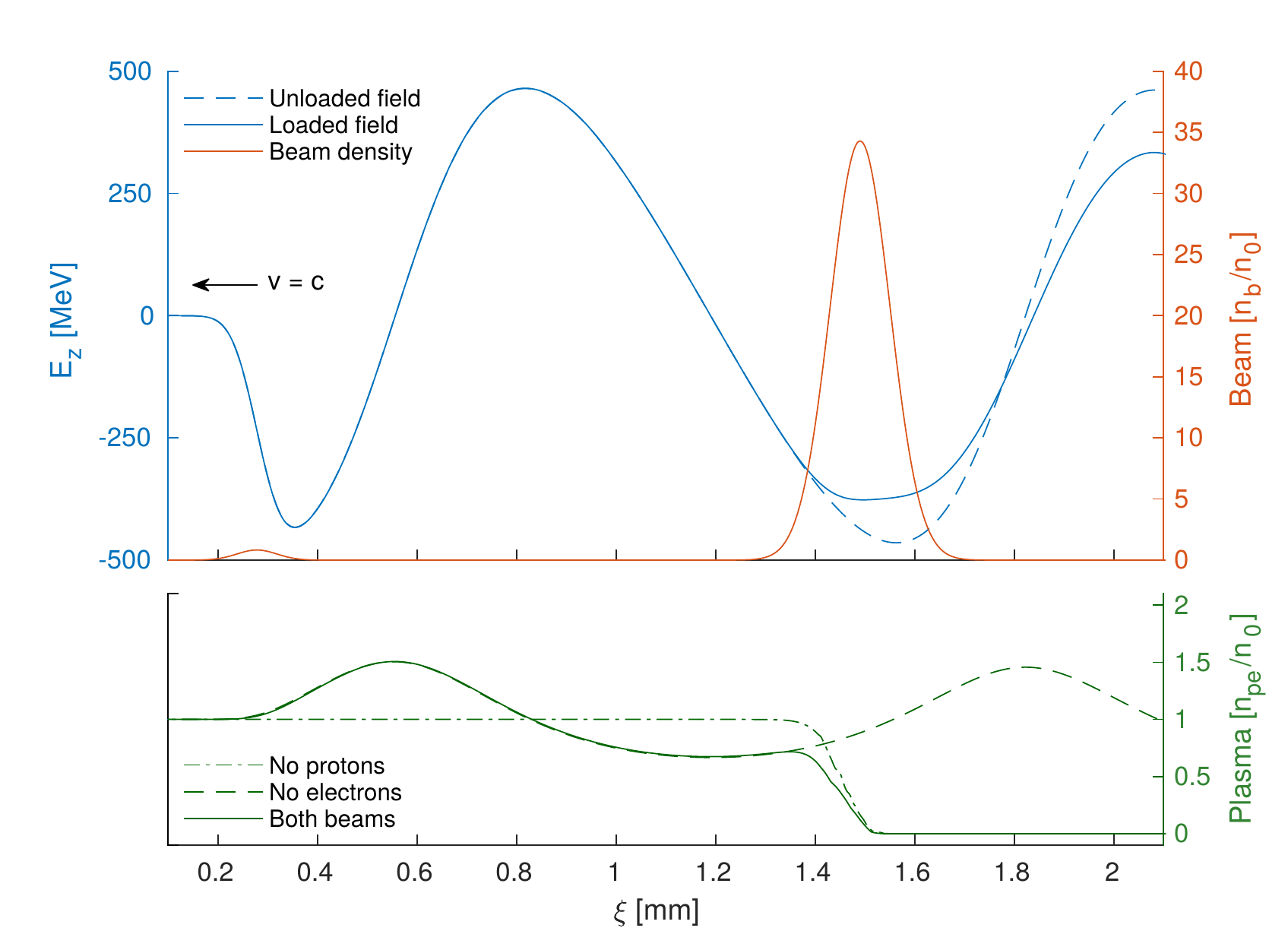}
    \caption{\label{Fig:BeamLoading} Top plot: Unloaded longitudinal electric field with no witness beam
        (dashed blue line) and loaded field (whole blue line) along the beam axis. The beam density along the axis for
        both beams are shown in red.
        Bottom plot: Plasma densities along the beam axis for a drive beam with no witness beam (dashed green line),
        witness beam with no drive beam (dash-dotted green line), and both beams present (continuous green line).
        The position in the simulation box $\xi = z - tc$, moving towards the left. The plots show the initial time
        step.}
\end{figure}

\begin{figure}[hbt]
    \includegraphics[width=\linewidth,trim={2mm 0mm 2mm 0mm},clip]{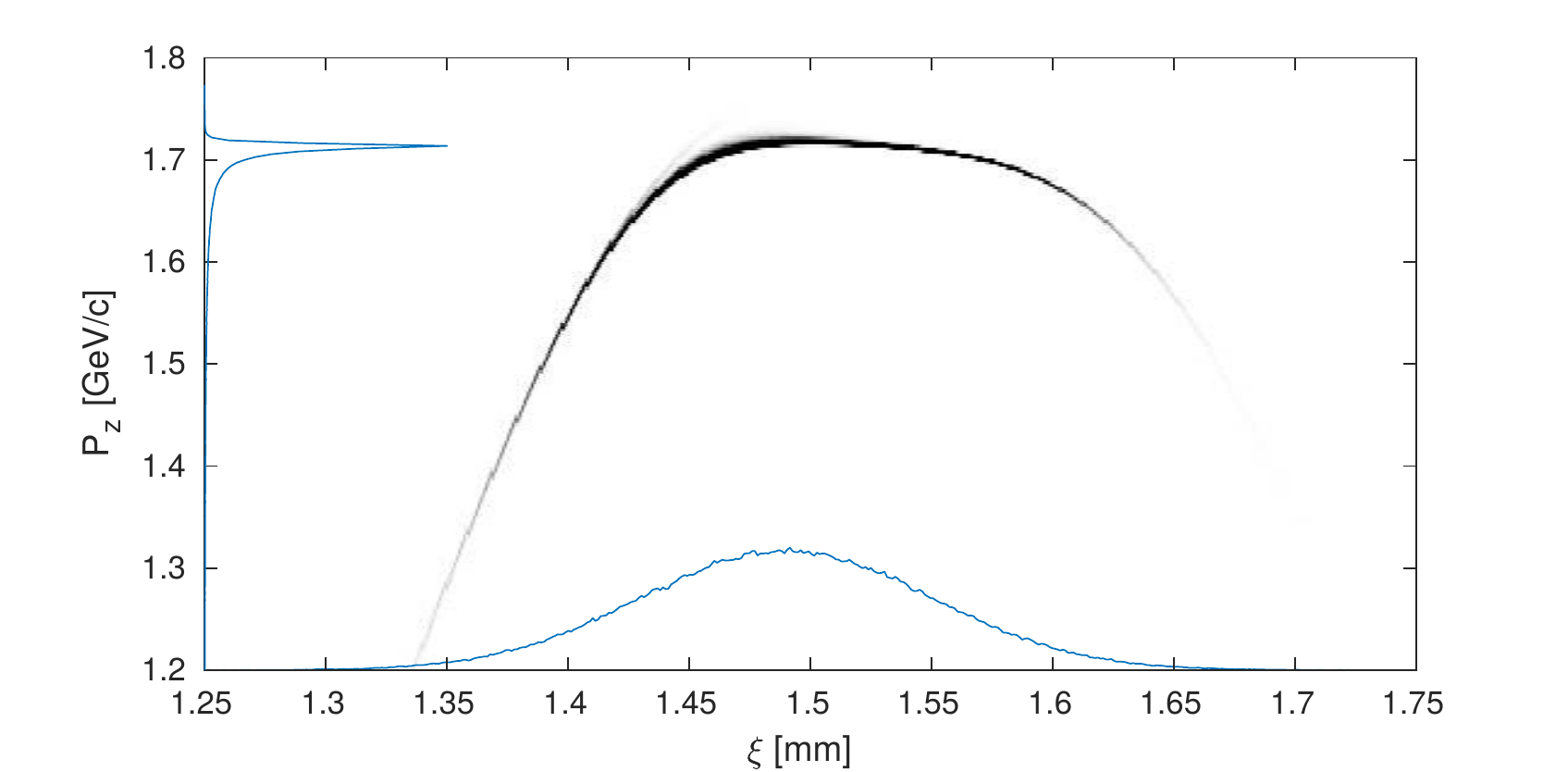}
    \caption{\label{Fig:BeamPS} Longitudinal phase space charge distribution of a $100\unit{pC}$, $60\unit{\mu m}$ long
        witness beam after $4\unit{m}$ of plasma. The mean momentum is $1.67\unit{GeV/c}$ with an RMS energy spread
        of $87\unit{MeV/c}$ ($5.2\%$) for the full beam.}
\end{figure}

\begin{figure*}[hbt]
    \includegraphics[width=0.495\linewidth,trim={2mm 0mm 2mm 0mm},clip]{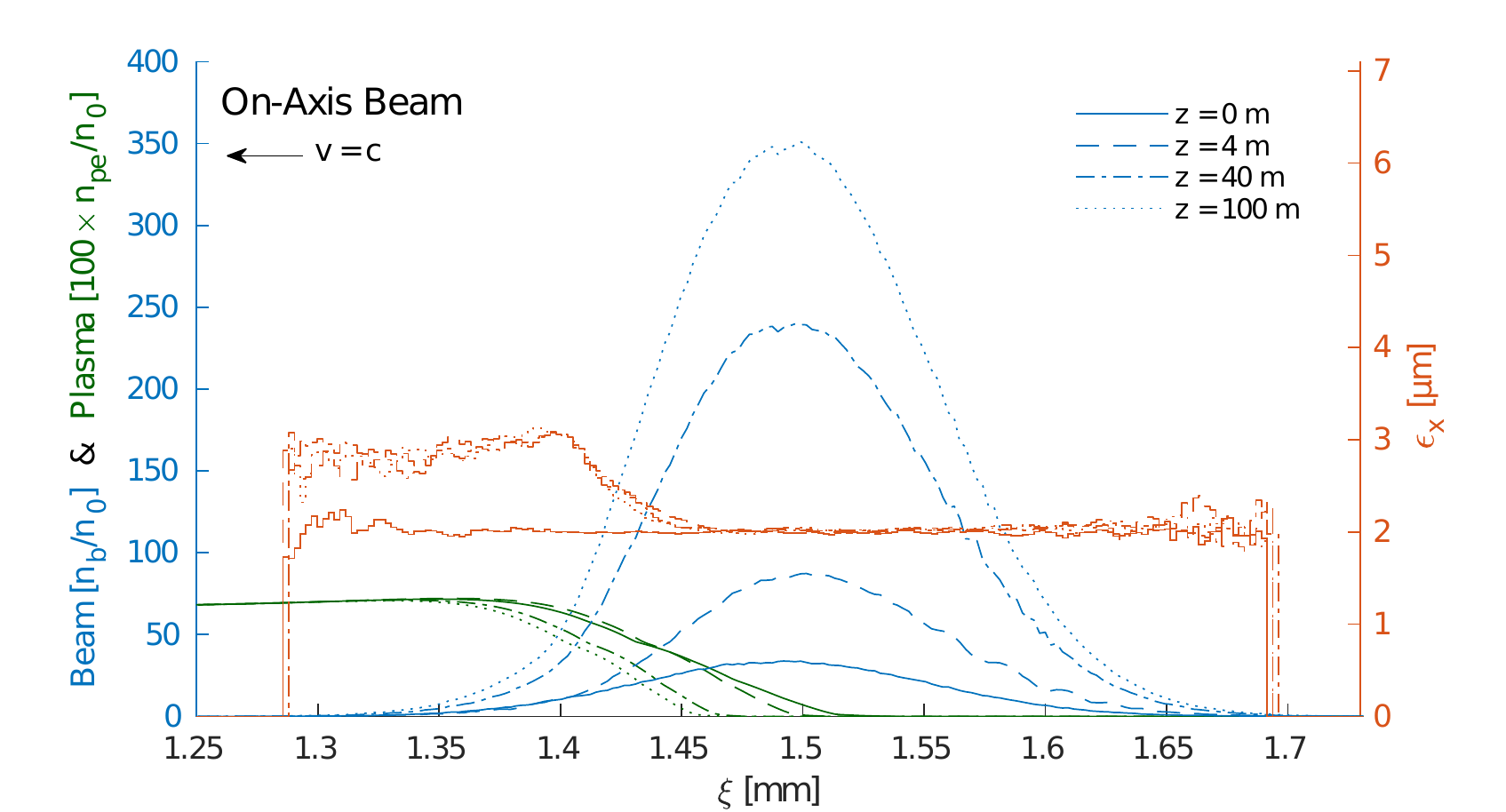}
    \includegraphics[width=0.495\linewidth,trim={2mm 0mm 2mm 0mm},clip]{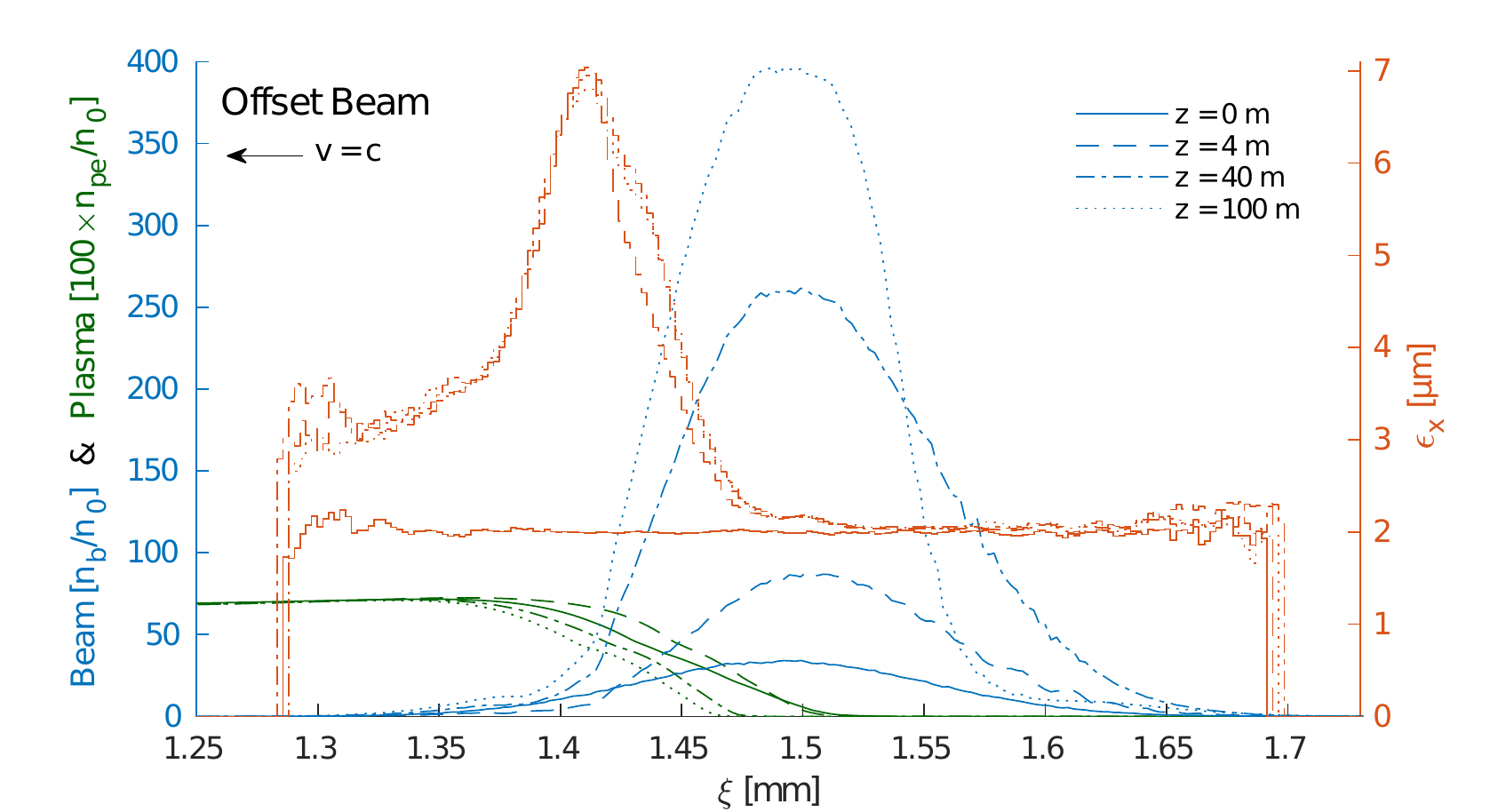}
    \caption{\label{Fig:BeamEmitt} Beam density in blue along the beam axis for an on-axis beam with respect to the
        drive beam axis (left), and an offset beam (right) with an offset of one $\sigma_{x,eb} = 5.24\unit{\mu m}$ in
        the x-plane \textendash at four different positions $z$ in the plasma stage. The red lines show a moving window
        calculation of transverse normalised emittance. The moving window calculation uses longitudinal slices of
        $l = 4\cdot\Delta\xi = 9.38\unit{\mu m}$ with a step of $\Delta\xi$. Only slices with more than $100$ macro
        particles have been included. The plasma density profile is included in green, and scaled up by a factor of
        $100$ to be visible. These simulations were run with an LHC energy drive beam of $7\unit{TeV}$}
\end{figure*}

\begin{figure*}[hbt]
    \includegraphics[width=\linewidth,trim={0mm 0mm 0mm 0mm},clip]{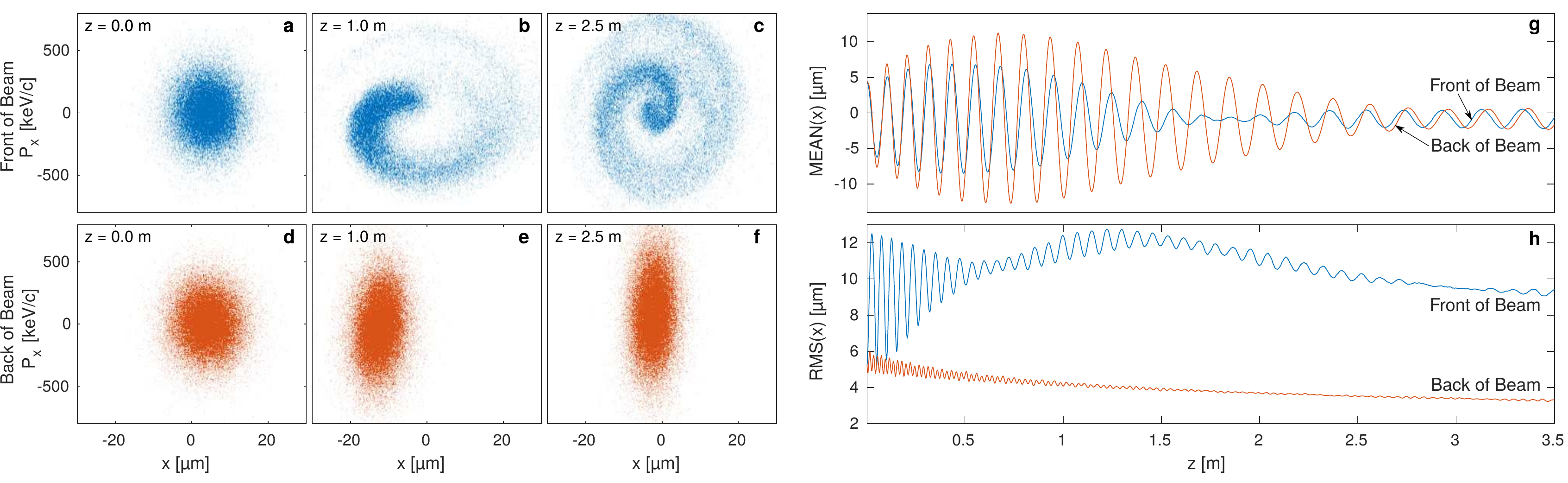}
    \caption{\label{Fig:BeamFilament} Plots \textbf{a} to \textbf{f} show the transverse phase space of the offset
        electron beam at different plasma positions. Plot \textbf{g} shows the macro particle mean position, and plot
        \textbf{h} their RMS spread. Plots \textbf{a}, \textbf{b} and \textbf{c}, as well as the blue lines in plots
        \textbf{g} and \textbf{h} represent particles not in the ion column (see Fig.~\ref{Fig:BeamEmitt}), with
        position $1.40\unit{\mu m} < \xi < 1.42\unit{\mu m}$. Plots \textbf{d}, \textbf{e} and \textbf{f}, and the red
        lines in plots \textbf{g} and \textbf{h} represent particles in the ion column with position
        $1.55\unit{\mu m} < \xi < 1.57\unit{\mu m}$.}
\end{figure*}

\section{Beam Loading}\label{S:BL}

Figure~\ref{Fig:BeamLoading} shows the results of QuickPIC simulations of the initial time step for the base case
parameters. The $E_{z}$-field generated by the proton drive beam is seen as the blue line, shown with and without the
electron beam present. With a proton beam density $n_{pb} \simeq n_{0}$, the wakefields are in the quasi-linear
regime~\cite{rosenzweig:2010}. The dashed green line in the lower part of Fig.~\ref{Fig:BeamLoading} shows that the
on-axis plasma density has a depletion to $67\%$, close to what we see in full scale reference simulations for AWAKE
Run~2~\cite{awake_collaboration:2016}.

The witness beam generates its own wakefield that loads the $E_{z}$-field generated by the drive beam. With an ideally
shaped electron beam charge profile it is possible to optimally load the field in such a way that the accelerating field
is constant along the beam~\cite{katsouleas:1987, tzoufras:2009}. Gaussian beams, as assumed in these studies, cannot
completely flatten the electric field in the tails of the charge distribution, and our base case beam therefore has a
tail in energy both at the front and the back of the beam, as illustrated in Fig.~\ref{Fig:BeamPS}. The bulk of the
beam, however, sees a relatively flat field.

The initial electron beam density is $n_{eb} \approx 35\cdot n_{0}$. This means that the witness beam's own wakefield
is in the fully non-linear regime, where the space charge force is sufficient to blow out all plasma electrons,
resulting in the formation of a pure ion column (see Fig.~\ref{Fig:BeamLoading}, bottom). This ion column, as is well
known~\cite{rosenzweig:1991}, provides a linear focusing force on the part the electron beam within the column, and
therefore prevents emittance growth for this part of the beam. This bubble and the focusing force is shown for our base
case in Fig.~\ref{Fig:PlasmaDenTWake}. The focusing field has a gradient of $20\unit{kT/m}$ near the beam axis,
corresponding to the matched field gradient.

\begin{figure*}[hbt]
    \begin{minipage}[t]{.48\textwidth}
        \includegraphics[width=\linewidth,trim={2mm 0mm 2mm 0mm},clip]{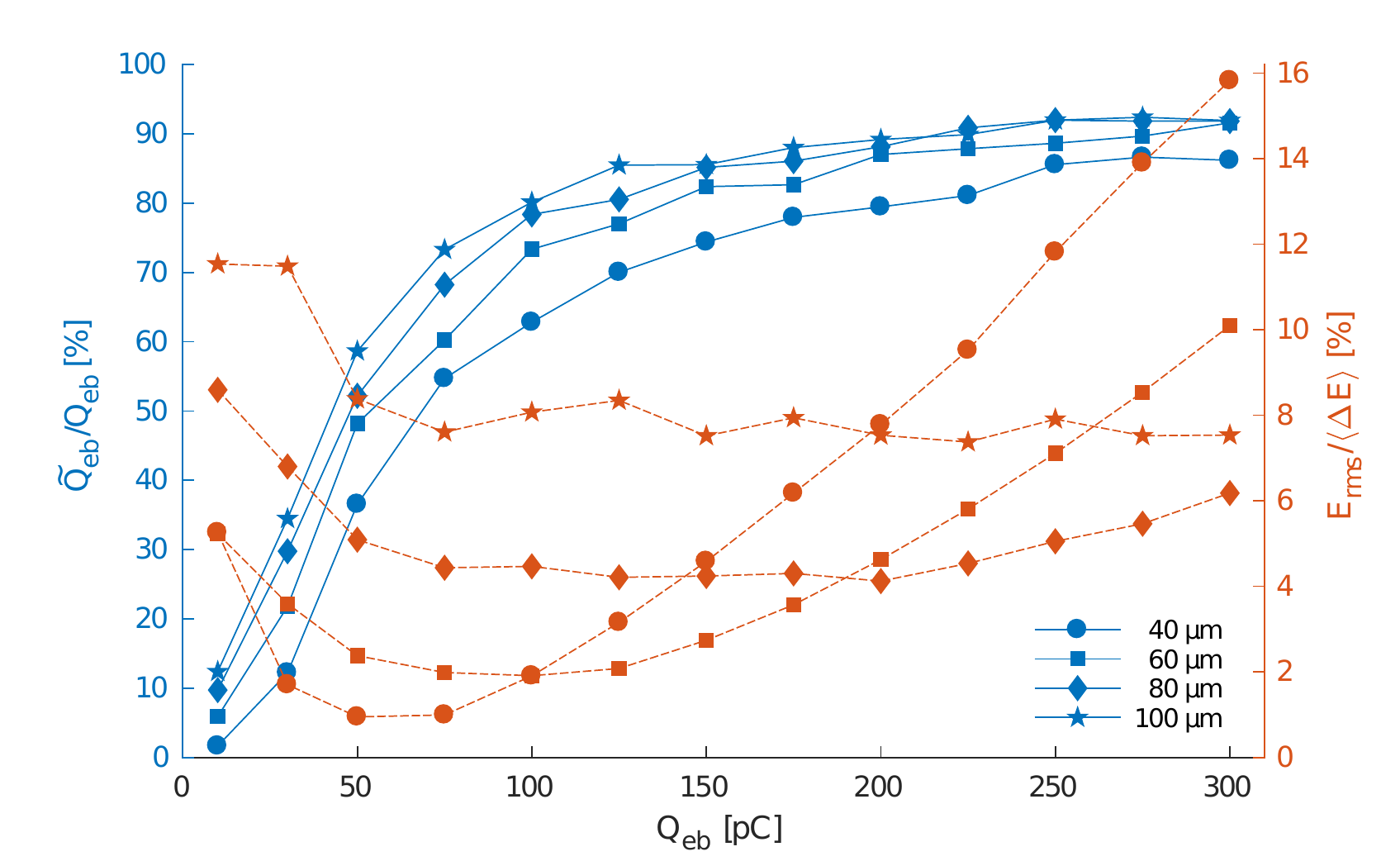}
        \caption{\label{Fig:BeamQ} Ratio of witness beam charge with emittance preserved, $\widetilde{Q}/Q$ (blue
            symbols, lines), as a function of initial beam charge, and relative energy spread of the accepted charge
            (red symbols, dashed lines), after $4\unit{m}$ of plasma and with an initial emittance
            $\epsilon_{\mathrm{N},0}=2\unit{\mu m}$. These are shown for four different $\sigma_{z}$ from
            $40\unit{\mu m}$ to $100\unit{\mu m}$. The detailed studies presented in beam loading section correspond to
            the square marked lines at $100\unit{pC}$.}
    \end{minipage}\hfill
    \begin{minipage}[t]{.48\textwidth}
        \includegraphics[width=\linewidth,trim={2mm 0mm 2mm 0mm},clip]{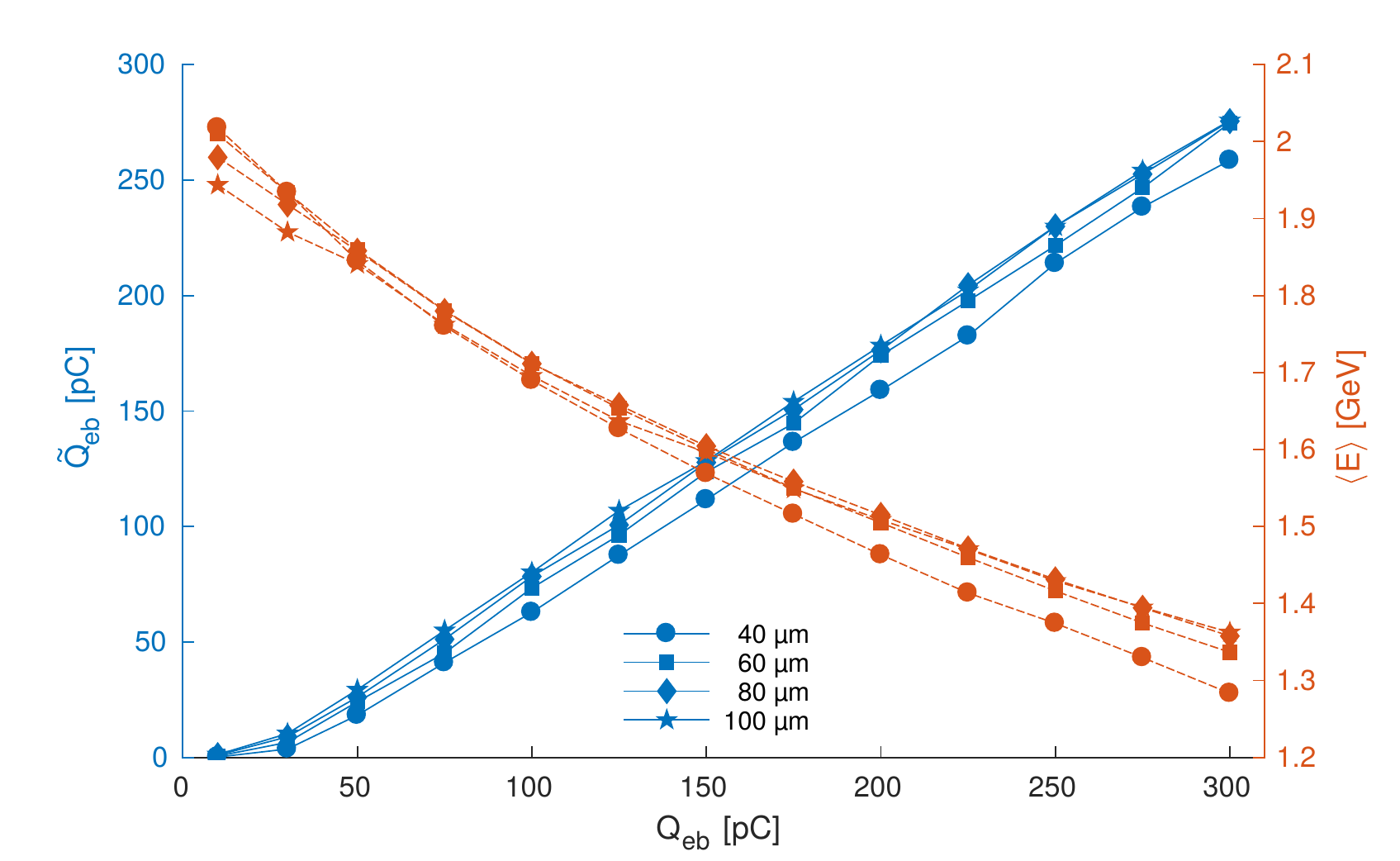}
        \caption{\label{Fig:BeamQAbs} Witness beam charge with emittance preserved, $\widetilde{Q}$ (blue symbols,
            lines), as a function of initial beam charge, and final energy (red symbols, dashed lines), after
            $4\unit{m}$ of plasma and with an initial emittance $\epsilon_{\mathrm{N},0}=2\unit{\mu m}$.}
    \end{minipage}
\end{figure*}

Figure~\ref{Fig:BeamEmitt} shows the slice emittance along the beam for the base case, sampled after propagating through
$0$, $4$, $40$ and $100\unit{m}$ of plasma. We define emittance of a slice as preserved if the growth is less than
$5\%$, and $\widetilde{Q}$ as the sum charge of the slices for which the emittance is preserved. Simulation results show
that $\widetilde{Q}/Q = 73\%$ of the electron beam longitudinal slices retain their initial emittance after the
propagation in the plasma. The total (projected) emittance of these slices combined is also preserved. Emittance growth
mainly occurs in the first few metres, and no significant emittance growth is observed after this for propagation
lengths up to $100\unit{m}$. The head of the beam does not benefit from the full ion column focusing, but since the
proton beam creates a quasi-linear wake, the emittance of the head of the beam still stabilises after some time. For the
$100\unit{m}$ simulation, the drive beam energy was increased to $7\unit{TeV}$ (LHC energy) to prevent de-phasing, as
de-phasing starts to become a significant effect for the SPS beam of $400\unit{GeV}$ after about $50\unit{m}$.

So far we have considered a witness beam injected on the axis of the proton beam. We now briefly examine the case of
injection of a witness beam with an offset with respect to the proton beam axis. Since the witness beam creates its own
plasma bubble, the emittance of the part of the beam inside that bubble is not affected by small transverse offsets of
the witness beam with respect to the proton beam axis. This is illustrated in Fig.~\ref{Fig:BeamEmitt}, right, for an
electron beam offset of one $\sigma_{x,eb}$. Emittance preservation for small offsets is an added benefit of this
accelerating regime, and may ease transverse injection tolerances. The head of the beam experiences a larger initial
emittance growth than for the on-axis case (compare Fig.~\ref{Fig:BeamEmitt}, left, to Fig.~\ref{Fig:BeamEmitt}, right).
However, also for the head of the beam the emittance growth ends after the first few metres.
Figure~\ref{Fig:BeamFilament}a-c show the phase space of the head of the electron beam after $0$, $1.0$ and
$2.5\unit{m}$, while Fig.~\ref{Fig:BeamFilament}d-f show the phase space of the trailing part of the beam. The
centroid oscillations of the head and the trailing part are shown in Fig.~\ref{Fig:BeamFilament}g. This effect of a
transverse offset is greater for larger offsets as the beam oscillates around the axis of the drive beam wakefield.

The transverse beam size within the bubble, where normalised emittance is preserved, follows the evolution given by
Eq.~\ref{EQ:Matched}; that is, evolves to stay matched. The on-axis density of the electron beam, as a result,
increases as its gamma factor increases and its transverse size decreases. This effect can be seen in Fig.
\ref{Fig:BeamFilament}h. This has the potential to cause overloading of the field. However, for the base case no
significant overloading is observed. Parameters can also be chosen in order to minimise this effect by slightly
underloading the wakefield at first, and let the high energy beam overload the wakefield at the end.

\section{Parameter Optimisation}\label{S:PO}

The beam loading and blow out properties of the electron beam depend on a large number of parameters, including
longitudinal profile, transverse profile, as well as relative phasing of the proton and the electron beams. We present a
limited parameter study aimed to guide beam parameter choices for AWAKE Run~2. For an electron beam to be externally
injected in AWAKE Run~2 (see Fig.~\ref{Fig:AWAKER2}) it is desired to maximise the energy gain, minimise the energy
spread, maximise the charge to be accelerated, and minimise the emittance growth~\cite{adli:2016}. In addition, the
beam length should be such that it is possible to generate and transport the beam using a compact electron
injector~\cite{adli:2016}. We investigate the interdependence of these parameters in simulation by varying the electron
beam length, its charge, and initial emittance. The results are quantified in terms of how much of the beam retains its
initial emittance. For these parameter scans we used a transverse grid cell size of $2.34\unit{\mu m}$, and let the
beams propagate through $4\unit{m}$ of plasma.

\begin{figure*}[hbt]
    \begin{minipage}[t]{.48\textwidth}
        \includegraphics[width=\linewidth,trim={2mm 0mm 2mm 0mm},clip]{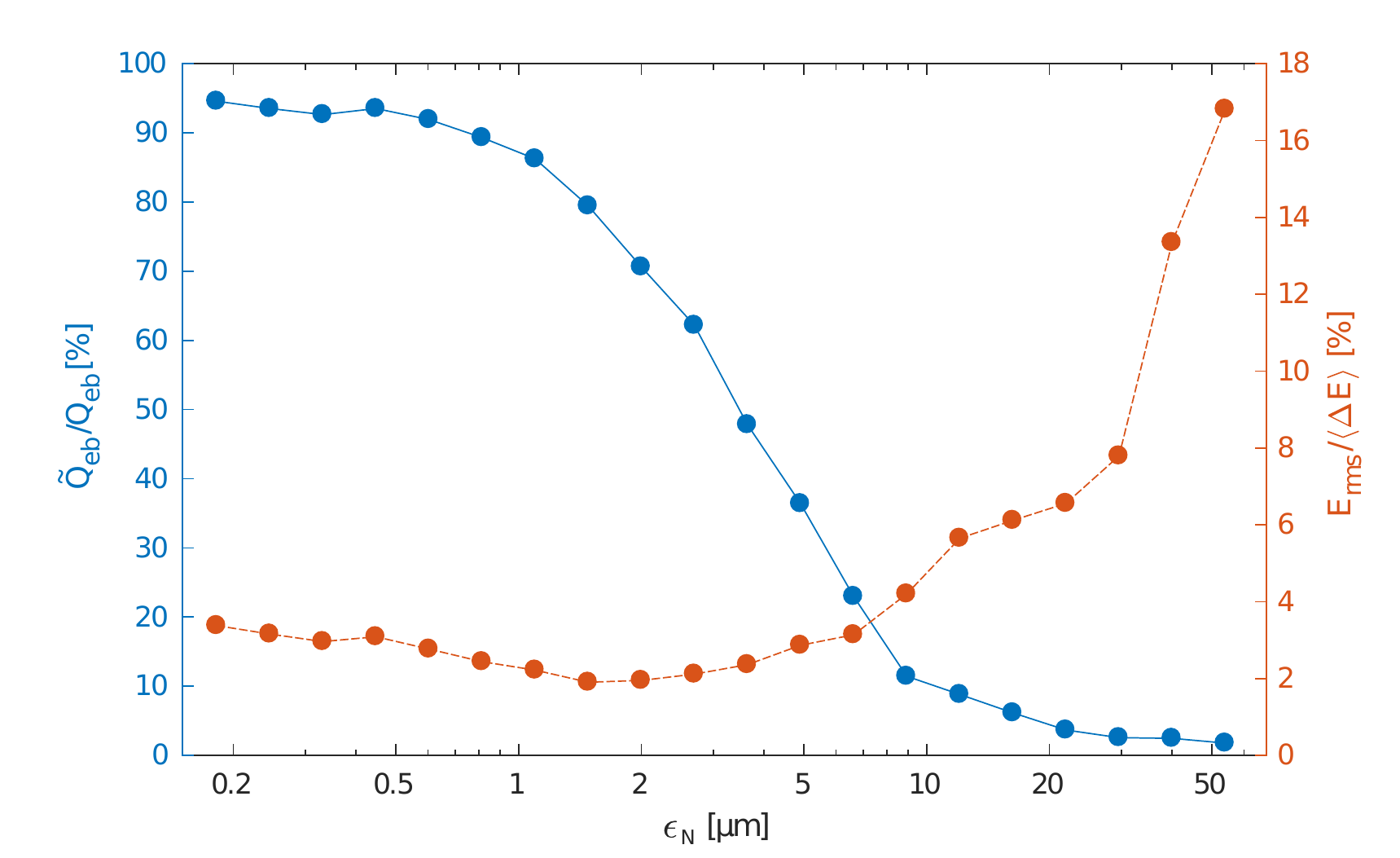}
        \caption{\label{Fig:BeamQEmit} Ratio of witness beam charge with emittance preserved, $\widetilde{Q}/Q$ (blue
            symbols, line), as a function of beam initial emittance (right), with relative energy spread of the accepted
            charge (red symbols, dashed line), after $4\unit{m}$ of plasma.}
    \end{minipage}\hfill
    \begin{minipage}[t]{.48\textwidth}
        \includegraphics[width=\linewidth,trim={2mm 0mm 2mm 0mm},clip]{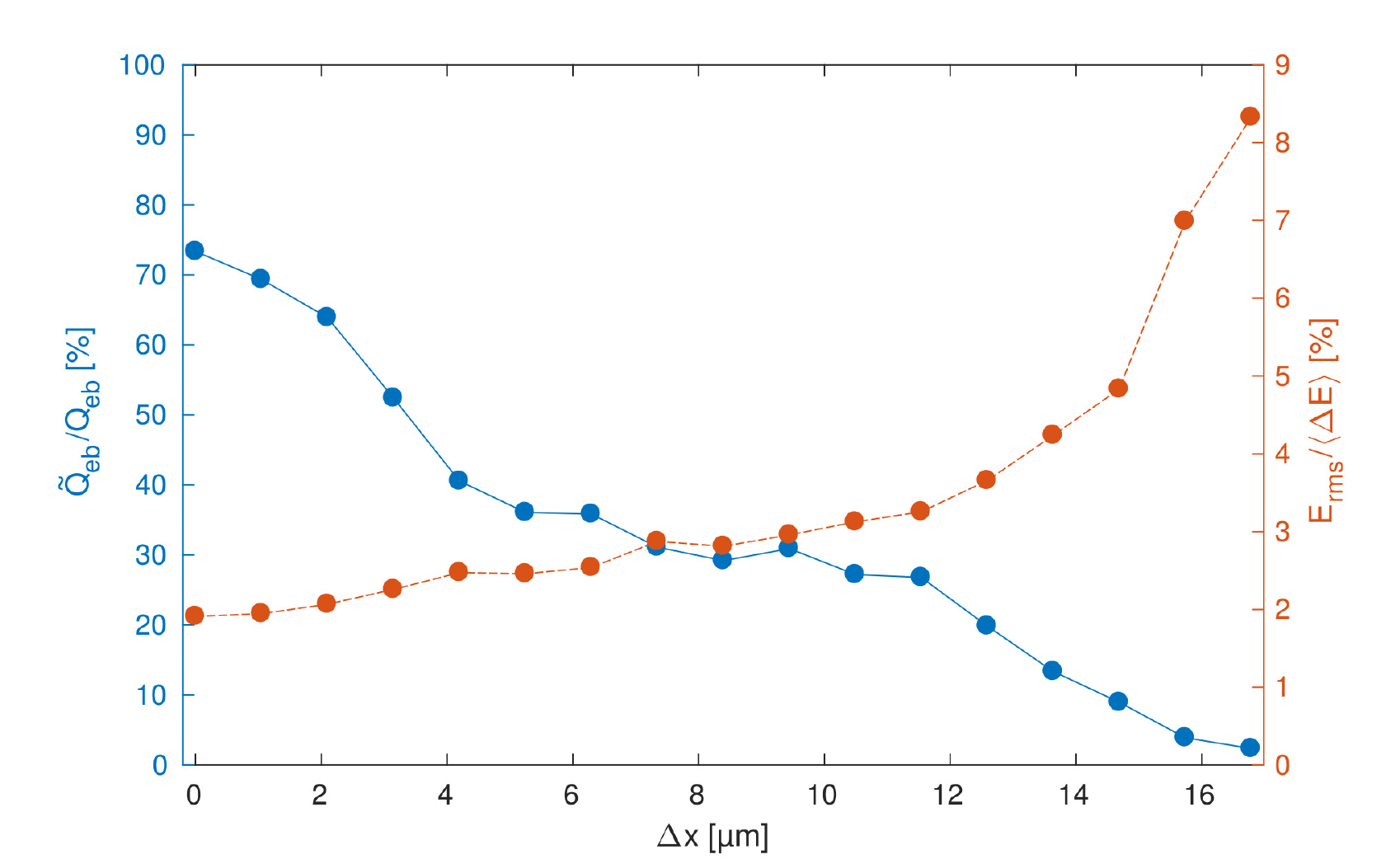}
        \caption{\label{Fig:BeamQOffset} Ratio of witness beam charge with emittance preserved, $\widetilde{Q}/Q$ (blue
            symbols, line), as a function of beam offset, with relative energy spread of the accepted charge (red
            symbols, dashed line), after $4\unit{m}$ of plasma.}
    \end{minipage}
\end{figure*}

Figure~\ref{Fig:BeamQ} shows the dependence of charge and energy spread on witness beam length and incoming charge. An
initial beam emittance of $2\unit{\mu m}$ was used. Therefore, we define the fractional charge $\widetilde{Q}$ as the
charge whose emittance remains smaller than $2.1\unit{\mu m}$. The beam charge ranges from $10\unit{pC}$ to
$300\unit{pC}$, and $\sigma_{z}$ ranges from $40\unit{\mu m}$ to $100\unit{\mu m}$. As can be seen from Fig.
\ref{Fig:BeamQ} (red curves), both the $40\unit{\mu m}$ and the $60\unit{\mu m}$ beams have a well defined minimum
energy spread with an initial beam charge $\approx 50\unit{pC}$ and $\approx 100\unit{pC}$, respectively. Lower beam
charges tend to underload the electric field, while higher beam charges tend to overload it. It is also clear that
longer beams with respect to the accelerating phase of the field, $\approx\lambda_{pe}/4$, do not optimally load the
wake, thus producing a larger spread in energy. The blue curves show the fraction of charge whose initial emittance is
preserved (the slice emittance for the base case is shown in Fig.~\ref{Fig:BeamEmitt}). As the witness beam charge
increases, the fraction of slices with preserved emittance increases \textendash as expected from an earlier onset of
the bubble formation \textendash and also increases in the bubble size~\cite{lu:2006-1, lu:2006}. We note here that
operation with $100\unit{pC}$ leads to a significantly larger charge (factor $\sim 4$) with emittance preserved, at the
expense of an increase of relative energy spread by a factor of two.

Figure~\ref{Fig:BeamQAbs} shows the dependence of mean energy gain on beam length and beam charge (red curves). The blue
curves show the amount of charge in the longitudinal slices where the emittance has been preserved, as a function of
beam length and beam charge. The results are weakly dependent on the electron beam length. As expected, a larger value
of $\widetilde{Q}$ corresponds to a lesser energy gain. No optimum is observed.

Figure~\ref{Fig:BeamQEmit} shows how the growth in emittance and energy spread varies with initial electron beam
emittance. In these simulations we adjusted the witness beam radius to maintain the matching condition at each
emittance. The smaller the initial emittance is, the better the emittance is preserved. There are two effects that lead
to emittance growth for high initial emittance beams: the transverse beam size may increase beyond the size of the
bubble, and the beam density may be reduced so much that the plasma electrons are no longer fully evacuated from the
bubble. Emittance values higher than a few micrometres lead to a significant increase in both emittance and energy
spread. 

Our base case showed some robustness to a small offset from the proton beam axis on the order of one $\sigma_{x,eb}$,
but with a reduction of the fraction of the beam which retains its initial emittance. Figure~\ref{Fig:BeamQOffset} shows
the correlation between this ratio for a range of offsets up to $16.8\unit{\mu m}$, corresponding to
$3.2\cdot\sigma_{x,eb}$. The effect on the head of the beam, shown in Figs.~\ref{Fig:BeamEmitt}
and~\ref{Fig:BeamFilament}, increases with larger offsets causing the part of the beam being defocused to extend
backwards to the point where the witness beam wakefields are no longer in the blow-out regime. At around
$3\cdot\sigma_{x,eb}$ emittance is no longer preserved at all.

The optimal working point will depend on the application and must be studied for each case and is, as illustrated in
this section, a trade off between beam length, beam charge and emittance preservation, as well as other parameters
like the plasma density.   

\section{Conclusion}\label{S:C}

We have devised a method to accelerate an electron witness beam to high energy with a low relative energy spread while
maintaining its incoming emittance in wakefields such that the accelerating structure is not void of plasma electrons.
This is the case for the AWAKE experiment in which the wakefields are driven by a train of proton bunches produce by
self-modulation of a long proton beam. This method is in principle applicable to all experiments operating in the
quasi-linear regime. Low relative energy spread and emittance preservation are achieved by choosing the electron beam
parameters to load the wakefields and evacuate the remaining plasma electrons from the accelerating structure.
 
Parameter studies indicate that for up to a few $100\unit{pC}$, about $70\%$ of the incoming beam charge is accelerated
for beam of lengths of $40-60\unit{\mu m}$. Such electron beams may be generated by an injector based on a standard
RF photo-emission gun~\cite{doebert:corr}.

In order to use manageable computer time for simulations, this study assumes a simplified case with respect to a
self-modulated proton beam, where the wake is driven by a single, short proton bunch producing similar wakefields.
However, the wakefields driven by a train of bunches evolve with the ramp of a real plasma and when entering the plasma.
Therefore, to be fully applicable to an experiment such as AWAKE, the study will have to be redone with more realistic
parameters. However, using loading of the wakefields and the pure plasma ion column fields to produce an accelerated
beam with low relative energy spread and emittance remains applicable.

\section{Acknowledgements}\label{Ack}

The simulations for this study have been performed using the open source version of QuickPIC released in early 2017
and owned by UCLA.

These numerical simulations have been made possible through access to the \emph{Abel} computing cluster in Oslo, Norway.
Abel is maintained by UNINETT Sigma2 AS and financed by the Research Council of Norway, the University of Bergen, the
University of Oslo, the University of Troms{\o} and the Norwegian University of Science and Technology. Project code:
nn9303k. Some of the simulations were also run on the student-maintained computing cluster \emph{Smaug} at the
University of Oslo, Department of Physics.

We gratefully acknowledge helpful discussions with Jorge Vieira from IST, Lisbon.
\vfill

\bibliography{Bibliography,Additional}
\end{document}